\documentclass[final,12pt]{elsarticle}

\usepackage{amsmath} 
\usepackage[displaymath]{lineno}
\usepackage{graphicx} 
\graphicspath{{fig/}}

\usepackage{float} 
\usepackage{subfigure}
\usepackage{color,soul}
\usepackage[version=4]{mhchem}
\usepackage[nodisplayskipstretch]{setspace}
\usepackage{booktabs} 
\usepackage[per-mode = symbol]{siunitx} 
\usepackage[margin=1in]{geometry}
\usepackage[section]{placeins} 
\usepackage{textcomp} 
\usepackage[flushleft]{threeparttable} 
\usepackage{tabularx}
\usepackage{array}
\newcolumntype{P}[1]{>{\raggedright\arraybackslash}p{#1}}
\newcolumntype{L}{>{\raggedright\arraybackslash}X}
\usepackage[labelsep=period,labelfont=bf]{caption} 
\usepackage{makecell}
\usepackage[hidelinks,bookmarksnumbered]{hyperref}

\makeatletter
\def\ps@pprintTitle{%
 \let\@oddhead\@empty
 \let\@evenhead\@empty
 \def\@oddfoot{}%
 \let\@evenfoot\@oddfoot}
\makeatother

\ExplSyntaxOn
\keys_define:nn { mhchem }
 {
  arrow-min-length .code:n =
   \cs_set:Npn \__mhchem_arrow_options_minLength:n { {#1} } 
 }
\ExplSyntaxOff
\mhchemoptions{arrow-min-length=1.5em}

\begin{document}

\begin{frontmatter}

\title{Dissolution kinetics of iron sulfide minerals in alkaline solutions}

\author[add1]{Zhanzhao Li\corref{cor1}}
\ead{zzl244@psu.edu}%
\author[add1]{Christopher A. Gorski}
\author[add2]{Aaron Thompson}
\author[add3]{Jeffrey R. Shallenberger}
\author[add4]{Gopakumar Kaladharan}
\author[add1]{Aleksandra Radlińska}

\cortext[cor1]{Corresponding author}
\address[add1]{Department of Civil and Environmental Engineering, The Pennsylvania State University, University Park, PA 16802, USA}
\address[add2]{Department of Crop and Soil Sciences, University of Georgia, Athens, Georgia 30602, USA}
\address[add3]{Materials Research Institute, The Pennsylvania State University, University Park, PA 16802, USA}
\address[add4]{USG Corporation, Libertyville, IL 60048, USA}

\begin{abstract}

Deleterious aggregate reactions induced by iron sulfide minerals, especially pyrrhotite and pyrite, have devastated concrete structures across many global regions. While these minerals have been extensively studied under acidic conditions, their behavior in alkaline environments, such as concrete, remains poorly understood. This study investigates the kinetics and mechanisms of iron sulfide dissolution at high pH (13--14). Results revealed that pyrrhotite dissolves orders of magnitude more rapidly than pyrite, with dissolution rates increasing with both pH and temperature. The type of alkali (potassium or sodium) in the solution was not found to affect the dissolution behavior. Kinetic modeling and experimental characterization indicated that the dissolution kinetics of pyrrhotite is controlled by a combination of chemical reactions (oxidation of iron and sulfur species) and diffusion (through an Fe(III)-(oxy)hydroxide layer). These findings provide practical insights into controlling dissolution and mitigating iron sulfide-induced damage in concrete.

\end{abstract}


\end{frontmatter}

\section{Introduction} 

Pyrrhotite (\ce{Fe_{1-x}S}) and pyrite (\ce{FeS2}), as the most abundant iron sulfide minerals in nature, are pivotal to a wide range of geological and industrial domains. Their reactivity and oxidation have received substantial attention for many decades, particularly in efforts to improve the efficiency of mineral processing and metal extraction and to mitigate the environmental impacts associated with acid mine drainage \cite{Belzile2004,Evangelou1995,Chandra2010}. However, most studies were carried out in acidic media \cite{Janzen2000,Nicholson1993,Descostes2004,Chirita2008,Chirita2014}; oxidation mechanisms under alkaline conditions, especially in high pH environments such as concrete (where pore solution typically exhibits a pH greater than \num{12.5} \cite{Vollpracht2016}), remain largely unexplored.

The presence of iron sulfide minerals in \textcolor{black}{concrete aggregates} has been found to be detrimental to concrete structures, causing durability issues for infrastructure in various regions around the world \cite{Chinchon1995,Rodrigues2012,Zhong2018,Han2018,Leemann2023,Jeyakaran2023}. \textcolor{black}{When exposed to oxygen and moisture, iron sulfides oxidize to form rust-like products (e.g., goethite, \ce{FeOOH}) as well as sulfuric acid \cite{Rodrigues2016a,Jana2020}. While the protons produced are typically buffered immediately by the highly alkaline concrete pore solution, the released sulfate ions can further react with the cement matrix to trigger internal sulfate attack and produce secondary products, such as ettringite (\ce{3CaO*Al2O3*3CaSO4*32H2O}, AFt), thaumasite (\ce{CaSiO3*CaCO3*CaSO4*15H2O}), and gypsum (\ce{CaSO4*2H2O}) \cite{Leemann2023}.} The formation of the rust-like products and secondary \textcolor{black}{sulfate minerals} induces internal stresses and expansion within concrete, resulting in cracking, spalling, and eventually crumbling of concrete structures. The economic and safety implications of such degradation, including increased maintenance and remediation costs and compromised structural integrity, underscore the urgent need for effective guidelines and mitigation strategies for the use of iron sulfide-bearing aggregates in concrete. The challenge in formulating these strategies, however, lies in the lack of a fundamental understanding of the complex, multi-mechanistic deterioration processes (i.e., \textcolor{black}{iron sulfide oxidation and subsequent internal sulfate attack}). 

Recent research efforts have been made to characterize concrete structures affected by iron sulfide reactions \cite{Schmidt2011,Rodrigues2012,Oliveira2014,Zhong2018,Leemann2023,Zhong2024} and develop evaluation methods to quantify the oxidation potential of iron sulfide-bearing aggregates \cite{Rodrigues2015,Rodrigues2016a,Geiss2019,El-Mosallamy2020a,El-Mosallamy2020,Cruz-Hernandez2020,Jeyakaran2023,Li2023,CastilloAraiza2023,CastilloAraiza2024,CastilloAraiza2024a,Guirguis2017,Ramos2016,Ojo2024}. Nevertheless, the concrete community has yet to reach a consensus on the use of iron sulfide-containing aggregates in concrete. For example, standard specifications for the use of aggregates in concrete, such as ASTM C33/C33M \cite{ASTMInternational2018a} and Canadian Standards Association (CSA) A23.1:19/A23.2:19 \cite{CanadianStandardsAssociation2019}, do not provide mandatory limits or requirements on acceptable amounts of iron sulfide minerals or total sulfur content in aggregates. While the European standard EN 12620:2013 \cite{BritishStandardsInstitution2013} has established a maximum total sulfur content of \qty{1}{\%} (or \qty{0.1}{\%} when pyrrhotite is present in the aggregate) for screening purposes, there has been debate about whether the specified limits are exaggerated and lead to unnecessary rejections of suitable aggregates \cite{Bentivegna2023,Li2023}. Beyond total sulfur content, numerous other factors can affect the rate and extent of iron sulfide reactions in concrete. These include \textcolor{black}{type of iron sulfides \cite{Rodrigues2012,Titon2024}}, concrete properties (e.g., pore solution chemistry, phase assemblage, and permeability) \textcolor{black}{\cite{Guirguis2018,El-Mosallamy2020a}}, and exposure conditions (temperature and moisture) \textcolor{black}{\cite{Leemann2024}}, to name a few. A more in-depth understanding of these influencing factors is thus required for further development of standard specifications and guidelines.

The reaction of iron sulfides in concrete starts with their dissolution into concrete pore solutions. The rate of this dissolution process can strongly influence the formation of expansive products and, consequently, the rate of concrete deterioration. Therefore, slowing down the dissolution rate could serve as an effective strategy to mitigate product formation and minimize the risk of damage. To achieve better control over the deleterious reactions, it is essential to establish a fundamental understanding of the dissolution kinetics of iron sulfide minerals in concrete pore solutions.

In light of the knowledge gaps mentioned above and the pressing need to address iron sulfide-induced deterioration in concrete, this study investigates the kinetics and mechanisms of iron sulfide dissolution in alkaline solutions. The main objectives are to: (1) evaluate the influence of different factors that can affect dissolution and subsequent reactions, including the \textcolor{black}{type of iron sulfides} (pyrrhotite or pyrite), alkali cation (potassium or sodium) and pH (13--14) of alkaline solutions, and temperature (23, 40, and \qty{60}{\celsius}); (2) determine the rate-controlling step of the dissolution process; and (3) identify and characterize the reaction products after dissolution. This investigation not only contributes to the body of knowledge in mineral processing, but also paves the way for practical solutions to enhance the durability and serviceability of concrete structures.

\section{Materials and methods}
\subsection{Mineral samples}

Natural pyrrhotite (Galax, Virginia, USA) and pyrite (Zacatecas, Mexico) samples used in this study were supplied by Ward's Science. The samples were crushed using a jaw crusher and sieved to \num{150}--\qty{300}{\um} (passing a No.~50 sieve and retained on a No.~100 sieve). This size fraction has been frequently used in mineral dissolution research \cite{Rimstidt2012,Kimball2010,Bagheri2022} and constitutes one of the major portions of fine aggregate in concrete \cite{ASTMInternational2018a,Munoz2021}.

After sieving, the samples were repeatedly rinsed with distilled water and washed in an ultrasonic bath until a clear, colorless supernatant solution was obtained. This allowed the removal of fine particles adhering to the crushed grain surface, avoiding anomalously high specific surface areas and thus initial dissolution rates \cite{McKibben1986,Bilenker2016}. The samples were then treated with \qty{10}{\%} HCl for \qty{2}{h} and rinsed with isopropanol to remove preexisting carbonate and oxide layers formed on the surfaces \cite{Li2016,Bilenker2016,Kocaman2016}. The cleaned samples were finally vacuum dried for \qty{24}{h} and stored in airtight bags flushed with \ce{N2} until the experiments to prevent oxidation. 

\begin{figure}[!htb]
	\centering
	\includegraphics[scale=0.9]{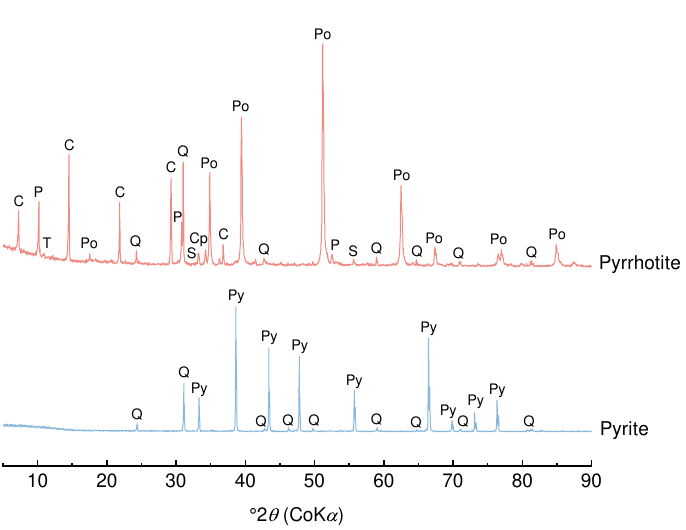}
	\caption{X-ray diffraction patterns of the initial, cleaned pyrrhotite and pyrite samples. Po: pyrrhotite (\ce{Fe_{1-x}S}, \#04-016-8469); Py: pyrite (\ce{FeS2}, \#04-014-6085); Cp: chalcopyrite (\ce{CuFeS2}, \#01-086-4137); S: sphalerite (\ce{ZnS}, \#01-071-5976);  Q: quartz (\ce{SiO2}, \#00-046-1045); C: chlorite (\ce{(Mg,Al)6(Si,Al)4O10(OH)8}, \#00-052-1044); P: phlogopite (\ce{K(Mg,Fe)3(Al,Fe)Si3O10(OH)2}, \#00-042-1437); T: talc (\ce{Mg3(Si2O5)2(OH)2}, \#01-073-0147).}
	\label{fig:XRD}
\end{figure}

Solid characterization by X-ray diffraction (XRD) confirmed that pyrrhotite or pyrite was the major component in the natural mineral samples, with minor impurities present ({\it cf.}~Fig.~\ref{fig:XRD}). The chemical composition of the samples was further determined through acid digestion and total sulfur content measurement. The results are listed in Table \ref{tab:chem_composition}. The purity of the natural pyrrhotite and pyrite samples was estimated as \qty{83.1}{\%} and \qty{94.5}{\%} by weight, respectively, based on the total sulfur content results after subtracting the contribution from sulfur-containing impurities.

\begin{table}[!htb]
\footnotesize
\centering
\caption{Composition of the initial, cleaned pyrrhotite and pyrite samples determined by acid digestion and total sulfur content measurement. Values are presented in wt\%.}
\label{tab:chem_composition}
	\begin{tabular}{llllllll}
		\toprule
         Sample & Al & Cu & Fe & Mg & Zn & S & Si (as \ce{SiO2}) \\ \midrule
        Pyrrhotite & 1.32 & 0.44 & 50.50 & 1.57 & 0.79 & 33.10 & 7.1 \\
        Pyrite & 0.10 & 0.10 & 45.40 & 0.00 & 0.04 & 50.40 & 7.8 \\
           \bottomrule
	\end{tabular}
\end{table}

Specific surface area was determined by the Brunauer–Emmett–Teller (BET) method using a Micromeritics\textsuperscript{\tiny\textregistered} TriStar II Plus surface area and porosity analyzer. The samples were degassed at \qty{70}{\degreeCelsius} under nitrogen for \qty{24}{h} before tests \cite{Multani2018}. The resulting BET surface areas for pyrrhotite and pyrite were \qty{0.74}{\square\m\per\g} and \qty{0.07}{\square\m\per\g}, respectively. The higher BET surface area for pyrrhotite compared to pyrite can be attributed to a greater extent of fractures of the pyrrhotite grains during milling and crushing \cite{Janzen2000,Belzile2004}.

\subsection{Dissolution experiments}

A series of batch reactor experiments were carried out to understand the dissolution behavior of iron sulfides in alkaline solutions. The details of the experimental parameters are summarized in Table \ref{tab:experiment}. For each experiment, \qty{0.2}{g} of pyrrhotite or pyrite sample was added to \qty{200}{mL} of solution stored in a high-density polyethylene container, resulting in a solution-to-solid ratio of \qty{1000}{mL/g}. Alkaline solutions were prepared using potassium hydroxide (\qty{\geq 85}{\%}, VWR Chemicals BDH\textsuperscript{\tiny\textregistered}) or sodium hydroxide (\qty{\geq 98}{\%}, Macron Fine Chemicals\textsuperscript{\texttrademark}) dissolved in ultrapure deionized water (\qty{18.2}{\Mohm\cm}). The concentrations between \qty{0.1}{M} and \qty{1.0}{M} were selected to represent the typical composition of pore solutions in cementitious systems \cite{Vollpracht2016,Bagheri2022}. \textcolor{black}{A headspace of air was left in each container to provide the necessary oxygen source for the dissolution and oxidation reactions. The containers were tightly closed to prevent loss of solution by evaporation and then placed in a convection temperature-controlled chamber.} Prior to introducing solid samples, the containers with the solutions were thermally equilibrated for over \qty{1}{h} at the temperatures specified in Table \ref{tab:experiment}. Agitation was performed using a rotary shaker (120 rpm) to ensure a continuously homogeneous solution.

\begin{table}[!htb]
\footnotesize
\centering
\caption{Summary of parameters for the dissolution experiments.}
\label{tab:experiment}
	\begin{tabular}{lll}
		\toprule
        Solid       & Solution         & Temperature (\unit{\degreeCelsius}) \\ \midrule
       
        Pyrrhotite & \qty{0.4}{M} KOH      & 23, 40, 60                     \\
                   & \qty{0.1}{M} or \qty{1.0}{M} KOH & 23                             \\
                 & \qty{0.4}{M} NaOH     & 23                           \\  
          Pyrite     & \qty{0.4}{M} KOH      & 23                             \\
        & \qty{0.4}{M} NaOH     & 23                                \\\bottomrule
	\end{tabular}
\end{table}

Each dissolution experiment was conducted for up to \qty{3}{h}. \textcolor{black}{At selected time intervals, the containers were briefly opened, and a \qty{1}{mL} solution aliquot was collected and filtered with \qty{0.2}{\um} nylon membrane.} The total volume of solution sampled for each experiment was minimal, with the change in the solution-to-solid ratio less than \qty{5}{\%}. Upon completion of the experiment, the solid residues were removed from the solution by vacuum filtration, rinsed with isopropanol, vacuum dried for \qty{24}{h}, and stored in airtight bags flushed with \ce{N2}. Analyses of the solution and solid samples were detailed in the following sections.

\subsection{Analysis of solutions}

pH was monitored before and after each dissolution experiment through acid titration, using phenolphthalein as the indicator immediately after filtration. The variation in pH was not significant (\num{< 0.1} units) due to the large solution-to-solid ratio. This ensured nearly constant pH conditions throughout the dissolution experiments.

The solution samples retrieved at specific time intervals were further diluted 10 times with \qty{2}{\%} \ce{HNO3} solution to avoid precipitates. Sulfur concentration was then measured by inductively coupled plasma atomic emission spectrometry (ICP-AES; Thermo Scientific\textsuperscript{\texttrademark} iCAP\textsuperscript{\texttrademark} 7400) as an indicator of dissolution. Three measurements were taken on each sample and their mean and standard deviation were reported. Although dissolved iron has frequently been analyzed in studies of iron sulfide dissolution under acidic conditions \cite{Nicholson1993,Belzile2004,Descostes2004,Chirita2014,Bilenker2016}, it was not considered in this work since it was often underestimated and below the detection limit due to the hydrolysis and precipitation of ferrous/ferric ions in the highly alkaline environments.

\subsection{Determination of the initial rate}

The kinetics of iron sulfide dissolution was assessed by determining the rate of sulfur release in the solutions using the initial rate method \cite{Rimstidt1993,Chirita2014}. This method has been widely adopted in mineral dissolution studies \cite{Kimball2010,Rimstidt2012,Chirita2014} and tends to be unbiased as an assumption of the reactant order or the rate law is not required \cite{McKibben1986,Janzen2000}. For each experiment, the concentration of sulfur ($M_{\mathrm{S}}$, \unit{\mol\per\L}) versus time ($t$, s) data were fit to a second-order polynomial:
\begin{equation}
    M_{\mathrm{S}}=a_2 +b_2 t+c_2 t^2 \label{eq:M_s}
\end{equation}
with its derivative given by:
\begin{equation}
    r_{\mathrm{S}}^{\prime}=\frac{d M_{\mathrm{S}}}{d t}=b_2+2 c_2 t
\end{equation}
or to a first-order polynomial: 
\begin{equation}
    M_{\mathrm{S}}=a_1 +b_1 t
\end{equation}
where the derivative is:
\begin{equation}
    r_{\mathrm{S}}^{\prime}=\frac{d M_{\mathrm{S}}}{d t}=b_1
\end{equation}
The coefficient $b$ of both of these fits gives the apparent rate of sulfur release ($r_{\mathrm{S}}^{\prime}$, \unit{\mol\per\L\per\s}) at $t=0$, which can be converted to the normalized rate ($r_{\mathrm{S}}$, \unit{\mol\per\square\m\per\s}) \cite{Kimball2010}:
\begin{equation}
r_{\mathrm{S}}=r_{\mathrm{S}}^{\prime} \frac{V}{A m}
\end{equation}
Here, $V$ is the volume of the solution (L). $A$ and $m$ are the specific surface area (\unit{\square\m\per\g}) and mass (g) of the solid, respectively.

\subsection{Characterization of solids}

The solid samples before and after dissolution were analyzed by scanning electron microscopy (SEM), energy-dispersive X-ray spectroscopy (EDS), X-ray photoelectron spectroscopy (XPS), and \ce{^57Fe} Mössbauer spectroscopy.

\subsubsection{Scanning electron microscopy/energy-dispersive X-ray spectroscopy}

SEM images were taken on a Thermo Scientific\textsuperscript{\texttrademark} Quanta 250 ESEM, with a \qty{5}{kV} accelerating voltage, to investigate the morphology of the samples before and after dissolution. The elemental composition of the samples was further measured at an accelerating voltage of \qty{15}{kV} by a high-resolution field-emission SEM (Thermo Scientific\textsuperscript{\texttrademark} Apreo S) equipped with an Oxford Ultim Max EDS detector. Oxford Instruments AZtec software was used to perform qualitative analyses of the EDS data.

\subsubsection{X-ray photoelectron spectroscopy}

To determine the chemical bonding and composition of the samples, XPS experiments were performed using a Physical Electronics VersaProbe III instrument equipped with a monochromatic Al K$\alpha$ X-ray source ($h\nu = \qty{1486.6}{eV}$) and a concentric hemispherical analyzer. The maximum analysis spot size of \qty{200}{\um} in diameter was used for each measurement. While XPS is ideal for characterizing the surface chemistry of materials after reaction, preliminary scans of the samples without post-processing (i.e., grinding) exhibited considerable uncertainty. This was due to the large grain size of the samples (\num{150}--\qty{300}{\um}), which was comparable to the maximum spot size, leading to analyses being performed on individual heterogeneous grains containing mineral impurities rather than on more homogeneous sample areas. To achieve reproducible results, the samples were ground with a mortar and pestle, which provided information more representative of the bulk characteristics. The grinding was performed in an anoxic glovebox and the samples were mounted on a special sample holder that could be transferred to the XPS chamber without exposure to air. 

Charge neutralization was performed using both low-energy electrons (\qty{< 5}{eV}) and argon ions. The binding energy axis was calibrated using sputter-cleaned copper (Cu($2p_{3/2}) = \qty{932.62}{eV}$, Cu($3p_{3/2}) = \qty{75.1}{eV}$) and gold foils (Au($4f_{7/2}) = \qty{83.96}{eV}$) \cite{Seah2001}. Peaks were charge referenced to \ce{CH$_x$} band in the C($1s$) spectra at \qty{284.8}{eV}. Quantification was done using instrumental relative sensitivity factors that account for the X-ray cross section and inelastic mean free path of the electrons. Acquired data were processed using CasaXPS software. Peaks were separated from the background by using a Shirley background. The fitting approaches for Fe(2$p_{3/2}$) and S(2$p$) spectra were adapted from Thomas {\it et al.}~\cite{Thomas1998}. Reference binding energies used for XPS spectral analyses are summarized in Table S1.

\subsubsection{\ce{^57Fe} Mössbauer spectroscopy}

Fe speciation of the solid samples was assessed by transmission \ce{^57Fe} Mössbauer spectroscopy under 295, 77, and \qty{5}{K} using a variable temperature He-cooled system with a 1024 channel detector. A \ce{^57Co} source (\qty{\sim 50}{mCi} or less) embedded in a Rh matrix was used at room temperature. The samples were mounted between two pieces of \qty{0.127}{mm} thickness Kapton tape and stored in airtight bags before being transferred to the spectrometer cryostat to minimize oxygen exposure. Velocity (i.e., $\gamma$-ray energy) was calibrated using an $\alpha$-Fe foil at \qty{295}{K} and all center shifts and peak positions are reported with respect to this standard. The transducer was operated in constant acceleration mode and folding was performed to achieve a flat background. Mössbauer spectra were analyzed using the Voigt-based fitting method of Rancourt and Ping \cite{Rancourt1991} in the Recoil\textsuperscript{\texttrademark} software, ISA Inc. Further details on spectral and phase analyses are provided in Section S2.

\section{Results}
\subsection{Effect of \textcolor{black}{iron sulfide type}}

\begin{figure}[!htb]
	\centering
	\includegraphics[scale=0.9]{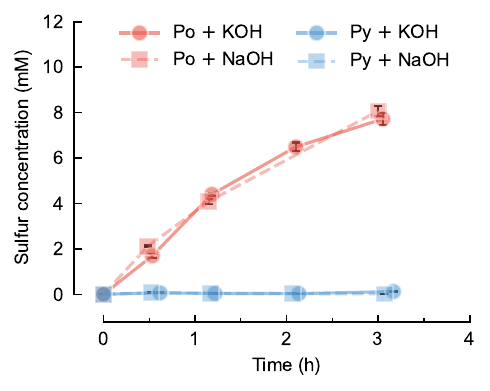}
	\caption{Concentration of sulfur released from the pyrrhotite (Po) and pyrite (Py) samples upon dissolution in \qty{0.4}{M} KOH and NaOH solutions. Error bars represent a \qty{95}{\%} confidence interval.}
	\label{fig:Sulfur_Mineral_Alkali}
\end{figure}

Table \ref{tab:rates} shows the experimental parameters and the rates determined from each dissolution experiment. Pyrrhotite and pyrite were added to a \qty{0.4}{M} KOH or NaOH solution to evaluate the effect of \textcolor{black}{iron sulfide type} on the dissolution behaviors. Representative relationships of the concentration of released sulfur as a function of time for the pyrrhotite and pyrite samples are displayed in Fig.~\ref{fig:Sulfur_Mineral_Alkali}.

\begin{table}[!htb]
\footnotesize
\centering
\caption{Dissolution rates of the pyrrhotite and pyrite samples in different solutions.}
\label{tab:rates}
\begin{threeparttable}
	\begin{tabular}{llllll}
		\toprule
         Solid      & Solution   & Temperature (\unit{\degreeCelsius}) & \multicolumn{2}{l}{pH$^a$} & Rate (\unit{\mol\per\square\m\per\s}) \\
        \cmidrule(l){4-5} & & & Initial & Final &         \\ \midrule
        Pyrrhotite & 0.4 M KOH & 23 & 13.6 & 13.6 & 1.65$\times 10^{-6}$ \\
         &  & 40 & 13.6 & 13.6 & 3.96$\times 10^{-6}$ \\
         &  & 60 & 13.6 & 13.6 & 1.17$\times 10^{-5}$ \\
         & 0.1 M KOH & 23 & 13.0 & 13.0 & 3.21$\times 10^{-7}$ \\
         & 1.0 M KOH & 23 & 14.0 & 14.0 & 2.79$\times 10^{-6}$ \\
         & 0.4 M NaOH & 23 & 13.6 & 13.6 & 1.54$\times 10^{-6}$ \\
        Pyrite & 0.4 M KOH & 23 & 13.6 & 13.6 & 2.87$\times 10^{-8}$ \\
         & 0.4 M NaOH & 23 & 13.6 & 13.6 & 1.93$\times 10^{-8}$ \\
        \bottomrule
	\end{tabular}
    \begin{tablenotes}
        \item $^a$ pH values were measured at room temperature.
	\end{tablenotes}
\end{threeparttable}
\end{table}

In a \qty{0.4}{M} KOH or NaOH solution, pyrrhotite exhibited a much higher rate and extent of sulfur release than pyrite, with differences in the rate reaching approximately three orders of magnitude. This underscores a greater dissolution rate and reactivity of pyrrhotite in alkaline environments. Based on the estimated purity of each mineral sample and the amount of solid used in each dissolution experiment, the maximum concentration of sulfur that could be released into solutions was calculated to be \qty{10.3}{mM} for pyrrhotite and \qty{15.8}{mM} for pyrite. A comparison between the sulfur concentration after 3-h dissolution for pyrrhotite (\qty{\sim 8}{mM}) and the theoretical maximum value (\qty{10.3}{mM}) shows that the majority of the pyrrhotite sample had dissolved (and reacted) under alkaline conditions at the end of the experiment. In contrast, the low concentration of sulfur released for pyrite (\qty{< 0.15}{mM}) indicates that most of the pyrite sample remained intact. These observations suggest that the dissolution of pyrrhotite is kinetically more favorable than that of pyrite, although more sulfur can be released from pyrite at a given mass theoretically. 

These findings align with previous dissolution studies conducted under acid conditions \cite{Steger1979,Thomas2000,Nicholson1993} and are further supported by recent characterizations of concrete affected by iron sulfide reactions \cite{Schmidt2011,Rodrigues2012}, which have revealed that pyrrhotite grains in sulfide-bearing aggregates react much faster than pyrite in alkaline concrete environments. The higher reactivity of pyrrhotite may be attributed to the lower stability of its crystal lattice due to the vacancy of iron atoms \cite{Nicholson1993}.

\subsection{Effect of alkali type}

Potassium and sodium are the most prevalent dissolved solutes in various aqueous environments, including the pore solutions in cementitious systems \cite{Dove1997,Vollpracht2016}. To investigate their influence on the dissolution of iron sulfides, experiments were conducted using \qty{0.4}{M} KOH or NaOH solutions. The results are presented in Table \ref{tab:rates} and Fig.~\ref{fig:Sulfur_Mineral_Alkali}. For pyrrhotite, the sulfur release rate in a \qty{0.4}{M} NaOH solution was 1.54$\times 10^{-6}$\,\unit{\mol\per\square\m\per\s}. This rate closely matches that observed in a \qty{0.4}{M} KOH solution ($1.65\times 10^{-6}$\,\unit{\mol\per\square\m\per\s}). Additionally, the final concentrations of the released sulfur in these solutions (\qty{7.7}{mM} in KOH and \qty{8.1}{mM} in NaOH) were also comparable. Similar observations were found in the dissolution of the pyrite sample, suggesting that the type of alkali ion exerts a negligible effect on the dissolution process. As a result, KOH was selected as the primary alkaline solution for subsequent investigations.

\subsection{Effect of pH}

As mentioned earlier, previous research on iron sulfide dissolution has primarily focused on acidic conditions; no study to our knowledge has examined its behavior in highly alkaline solutions. This work presents the first endeavor to investigate the effect of high pH on the dissolution kinetics of pyrrhotite. The initial pH was adjusted by varying the KOH concentration from \qty{0.1}{M} to \qty{1.0}{M}, resulting in pH values ranging from 13.0 to 14.0. Due to the large solution-to-solid ratio adopted in each experiment, fluctuations in pH were minimal, allowing a nearly constant pH throughout the experiment.

\begin{figure}[!htb]
	\centering
	\includegraphics[scale=0.9]{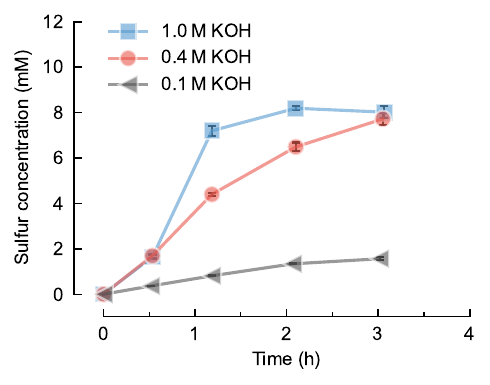}
	\caption{Concentration of sulfur released from the pyrrhotite samples upon dissolution in different KOH solutions. Error bars represent a \qty{95}{\%} confidence interval.}
	\label{fig:Sulfur_pH}
\end{figure}

Sulfur release rates at various pH levels derived from the dissolution experiments are provided in Table \ref{tab:rates}. The results demonstrate a notable increase in pyrrhotite dissolution rates with increasing pH. Specifically, the rates increased by 5.1- and 8.7-fold when the pH was raised from 13.0 to 13.6 and further to 14.0, respectively. Fig.~\ref{fig:Sulfur_pH} further illustrates the extent of sulfur release at different pH. The amounts of released sulfur at the completion of the experiments in \qty{0.4}{M} and \qty{1.0}{M} KOH solutions were approximately \qty{400}{\%} higher than that in a \qty{0.1}{M} KOH solution, highlighting the pH dependence of pyrrhotite dissolution under highly alkaline conditions. Further discussions of these results are provided in Sections \ref{sec:pH_dependence} and \ref{sec:implications}.

\subsection{Effect of temperature} \label{sec:temperature}

\begin{figure}[!htb]
	\centering
	\includegraphics[scale=0.9]{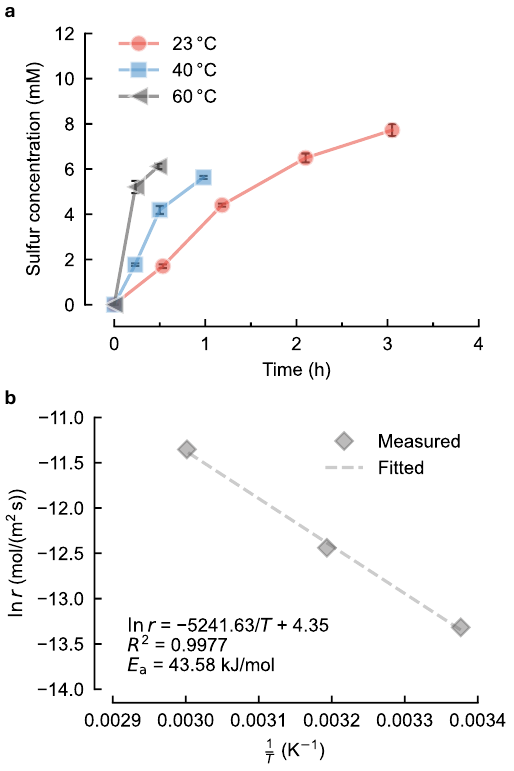}
	\caption{(a) Concentration of sulfur released from the pyrrhotite samples upon dissolution in \qty{0.4}{M} KOH solutions at different temperatures. Error bars represent a \qty{95}{\%} confidence interval. (b) Arrhenius plot for the apparent activation energy ($E_\text{a}$) for pyrrhotite dissolution. $r$, rate; $T$, temperature; $R^2$, coefficient of determination.}
	\label{fig:Sulfur_Temp}
\end{figure}

The influence of temperature on the dissolution was studied on the pyrrhotite sample in a \qty{0.4}{M} KOH solution ($\text{pH} = 13.6$) at 23, 40, and \qty{60}{\celsius}. As detailed in Table \ref{tab:rates} and Fig.~\ref{fig:Sulfur_Temp}a, an increase in temperature from \qty{23}{\celsius} to \qty{40}{\celsius} and \qty{60}{\celsius} resulted in 2.4- and 7.1-fold accelerations in the sulfur release rate, respectively. This indicates a strong temperature dependence of pyrrhotite dissolution. The effect of temperature on the dissolution rate followed the Arrhenius law \cite{Lasaga1984}, where the apparent activation energy was derived to be \qty{43.58}{\kJ\per\mol} (Fig.~\ref{fig:Sulfur_Temp}b). The value of activation energy offers insights into the kinetic control regimes governing the dissolution process. Activation energies below \qty{20}{\kJ\per\mol} are generally indicative of a diffusion-controlled mechanism, while those between 20 and \qty{80}{\kJ\per\mol} suggest that the dissolution is controlled by a mix of chemical reactions and diffusion \cite{Lasaga1998,DeGiudici2005}. As such, it can be inferred that the dissolution of pyrrhotite at pH 13.6 is likely under mixed control. Further investigations of the kinetic control regime for pyrrhotite dissolution are provided in Section \ref{sec:kinetic_control_regime}. The activation energy derived in this study appears to be consistent with those reported under more acidic conditions, as summarized in Table \ref{tab:activation_energy}.

\begin{table}[!htb]
\footnotesize
\centering
\caption{Activation Energies ($E_\text{a}$) for pyrrhotite dissolution.}
\label{tab:activation_energy}
	\begin{tabular}{llll}
		\toprule
         $E_\text{a}$ (\unit{\kJ\per\mol}) & pH & Temperature (\unit{\degreeCelsius}) & Reference \\ \midrule
        58.1 & 2 & 10--33 & Nicholson and Scharer \cite{Nicholson1993} \\
        47.7--62.5 & 2.5 & 25--45 & Janzen {\it et al.}~\cite{Janzen2000} \\
        41.6 & 3 & 25--45 & Chiriţǎ {\it et al.}~\cite{Chirita2008} \\
        40.3 & 3 & 4--35 & Romano \cite{Romano2012} \\
        52.4 & 4 & 10--33 & Nicholson and Scharer \cite{Nicholson1993} \\
        100.4 & 6 & 10--33 & Nicholson and Scharer \cite{Nicholson1993} \\
        65.3, 71.9 & 10.5 & 25--45 & Lehmann {\it et al.}~\cite{Lehmann2000} \\
        43.6 & 13.6 & 23--60 & This study \\
           \bottomrule
	\end{tabular}
\end{table}

\subsection{Determination of kinetic control regime} \label{sec:kinetic_control_regime}

Dissolution of iron sulfides involves multiple processes and reactions. Understanding the rate-controlling step is essential to reveal the underlying dissolution mechanisms. The shrinking core model has been widely used to describe the kinetics of mineral dissolution processes \cite{Levenspiel1999,Faraji2022}. In this model, three main controlling mechanisms are considered: (1) diffusion of the reactant through the thin liquid film surrounding the particle; (2) diffusion of the reactant through the solid product layer; and (3) reaction of the reactant with the solid on the surface of the unreacted core. In the case of spherical particles with unchanging size, the mathematical expressions are given as follows:
\begin{align}
    &\text{Liquid film diffusion control:~} \frac{t}{\tau_{F}}=X  \\
    &\text{Product layer diffusion control:~} \frac{t}{\tau_{P}}=1-3(1-X)^{2 / 3}+2(1-X) \\
    &\text{Chemical reaction control:~} \frac{t}{\tau_{R}}=1-(1-X)^{1 / 3} 
\end{align}
Here, $X$ is the fraction of the dissolved solid particles at the reaction time $t$, calculated by dividing the measured concentration of released sulfur by the theoretical maximum sulfur concentration. $\tau_{F}$, $\tau_{P}$, and $\tau_{R}$ are the time required for complete dissolution ($X=1$) under each controlling mechanism.

To determine the rate-controlling mechanism, these formulas need to be tested against experimental data, and the formula that best describes the data (often indicated by correlation coefficients $R$) will indicate the prevalent mechanism. While this approach has been frequently adopted, it becomes less effective when the performance of the formulas is very close and when more than one mechanism may govern the dissolution process (as indicated in Section \ref{sec:temperature}). To overcome these challenges, a more general expression that considers multiple mechanisms (i.e., liquid film diffusion control, product layer diffusion control, and chemical reaction control) was used in this study \cite{Levenspiel1999}:
\begin{equation}
    t= \tau_{F} X+\tau_{P}\left[1-3(1-X)^{2 / 3}+2(1-X)\right] +\tau_{R}\left[1-(1-X)^{1 / 3}\right] \label{eq:general_scm}
\end{equation}
The constants ($\tau_{F}$, $\tau_{P}$, and $\tau_{R}$) in Eq.~\eqref{eq:general_scm} were determined through a constrained least square optimization by minimizing the objective function $\phi$, subject to the constraints $\tau_{F}$, $\tau_{P}$, and $\tau_{R}>0$ \cite{Nazemi2011,Golmohammadzadeh2017}:
\begin{equation}
    \varphi= \sum_i\left(\tau_{F} X_i+\tau_{P}\left[1-3\left(1-X_i\right)^{2 / 3}+2\left(1-X_i\right)\right]\right. \left.+\tau_{R}\left[1-\left(1-X_i\right)^{1 / 3}\right]-t_i\right)^2
\end{equation}
where the subscript $i$ indicates the number of experimental data points.

Table \ref{tab:general_scm} shows the results obtained from fitting the experimental data to Eq.~\eqref{eq:general_scm} for pyrrhotite dissolution. The time contributions of the controlling mechanisms in the dissolution process are indicated by the values of $\tau_{F}$, $\tau_{P}$, and $\tau_{R}$. A non-zero value means the involvement of the controlling mechanism, whereas a zero value suggests no contribution from that mechanism to the dissolution kinetics. The sum of these constants ($\tau_{F}+\tau_{P}+\tau_{R}$) represents the total time required for the complete dissolution of pyrrhotite under the given experimental conditions.

\begin{table}[!htb]
\footnotesize
\centering
\caption{Parameters obtained by fitting experimental data to Eq.~\eqref{eq:general_scm} for pyrrhotite dissolution.}
\label{tab:general_scm}
\begin{threeparttable}
	\begin{tabular}{llllll}
		\toprule
         Solution & Temperature (\unit{\degreeCelsius}) & $\tau_{F}$ (h) & $\tau_{P}$ (h) & $\tau_{R}$ (h) & $R$ \\ \midrule
        0.1\,M KOH & 23 & 0.00 & 91.29 & 36.45 & 0.9920 \\
        0.4\,M KOH & 23 & 1.13 & 3.56 & 2.50 & 0.9973 \\
         & 40 & 0.39 & 3.73 & 0.84 & 0.9912 \\
         & 60 & 0.00 & 2.50 & 0.00 & 0.9760 \\
        1.0\,M KOH & 23 & 0.00 & 0.61 & 3.88 & 0.9576 \\
           \bottomrule
	\end{tabular}
     \begin{tablenotes}
         \item \textit{Note}: $\tau_{F}$, $\tau_{P}$, and $\tau_{R}$ represent the time required for complete dissolution under liquid film diffusion control, product layer diffusion control, and chemical reaction control, respectively, while $R$ is the correlation coefficient.
 	\end{tablenotes}
\end{threeparttable}
\end{table}

Pyrrhotite dissolution in a \qty{0.4}{M} KOH solution at \qty{23}{\degreeCelsius} was found to involve all three mechanisms: liquid film diffusion control ($\tau_{F} = \qty{1.13}{h}$), product layer diffusion control ($\tau_{P}=\qty{3.56}{h}$), and chemical reaction control ($\tau_{R} =\qty{2.50}{h}$). As temperature increased, the time contributions from liquid film diffusion control and chemical reaction control were reduced. At \qty{60}{\degreeCelsius}, the controlling mechanism shifted entirely towards product layer diffusion ($\tau_{P}=\qty{2.50}{h}$), with no significant roles of either liquid film diffusion or chemical reaction controls. The reduced contribution of chemical reactions in the dissolution process aligns with the commonly observed trend where diffusion mechanisms become more dominant over chemical reactions at elevated temperatures \cite{Brantley2008}. These findings are supported by the apparent activation energy derived in Section \ref{sec:temperature} (\qty{43.58}{\kJ\per\mol}), which suggests a mix of controlling mechanisms. Additionally, the reduction in total time required for complete dissolution with increasing temperatures, estimated by the sum of the constants, agrees with the faster dissolution rates measured at higher temperatures ({\it cf.}~Table \ref{tab:rates}).

In a \qty{0.1}{M} KOH solution, the dissolution of pyrrhotite was estimated to be much slower, with the total time required for complete dissolution calculated to be \qty{127.75}{h} (Table \ref{tab:general_scm}). The dissolution was predominately controlled by product layer diffusion, followed by chemical reaction mechanism. As the concentration of KOH increased to \qty{0.4}{M} and \qty{1.0}{M}, the total dissolution time decreased to \qty{7.19}{h} and \qty{4.49}{h}, respectively. It is also interesting to note that the contribution of product layer diffusion control diminished with higher KOH concentrations, while chemical reaction control became more pronounced. This shift is likely due to the increased ionic strength of the solution, which weakens the diffusion barrier formed by the product layer on the surface of the solid particles.

\subsection{Characterization of surface texture} \label{sec:surface}

\begin{figure}[!htb]
	\centering
	\includegraphics[scale=0.9]{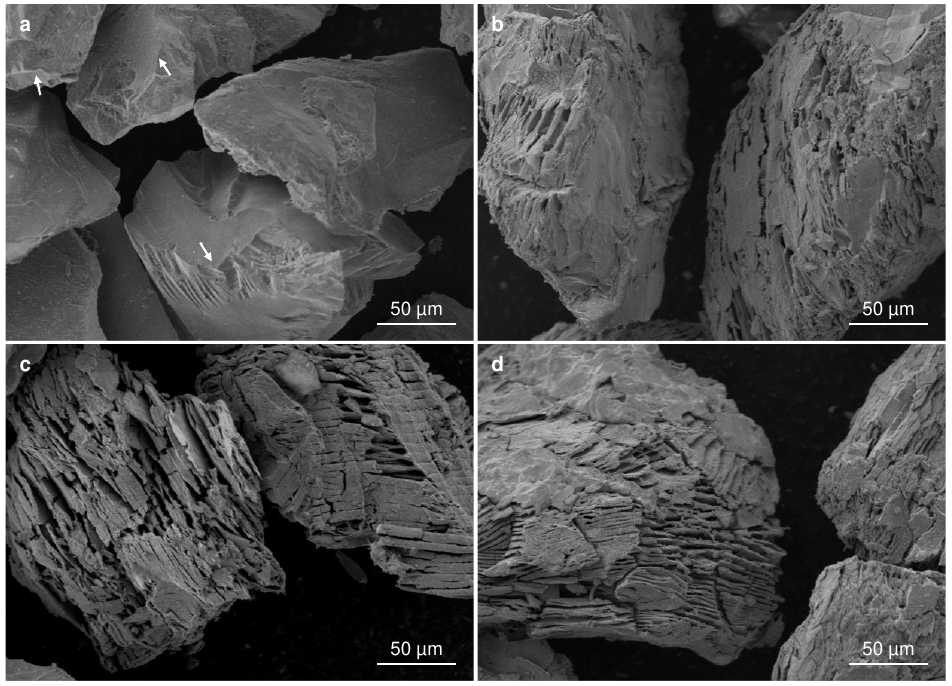}
	\caption{Scanning electron microscopy images of (a) the initial, cleaned pyrrhotite sample and the reacted pyrrhotite samples after 3-h dissolution in (b) 0.1\,M, (c) 0.4\,M, and (d) 1.0\,M KOH solutions. The arrows in (a) mark the regions with extensive striations and greater surface roughness, where preferential dissolution at higher rates could occur and lead to the formation of lamellar structures shown in (b), (c), and (d).}
	\label{fig:SEM_Dissolution}
\end{figure}

Fig.~\ref{fig:SEM_Dissolution} presents the surface morphology of pyrrhotite before and after the dissolution experiments in different concentrations of KOH solutions. The initial, cleaned pyrrhotite sample before experiments displayed smooth surfaces with sharp edges and no visible fine particles adhering (Fig.~\ref{fig:SEM_Dissolution}a). This demonstrates the effectiveness of the cleaning process, which is crucial for avoiding anomalously high dissolution rates and unexpected dissolution behaviors \cite{McKibben1986,Janzen2000}.

After 3-h dissolution in KOH solutions, the pyrrhotite samples exhibited distinct changes in their surface morphology. The sample exposed to \qty{0.1}{M} KOH appeared rather rough surfaces with some etch pits and emerging lamellar (layered, sheet-like) structures (Fig.~\ref{fig:SEM_Dissolution}b), while those treated with \qty{0.4}{M} or \qty{1.0}{M} KOH displayed even more pronounced lamellar formation (Fig.~\ref{fig:SEM_Dissolution}c,d). These morphological transformations correlate with the significant differences in rate (Table \ref{tab:rates}) and extent (Fig.~\ref{fig:Sulfur_pH}) of dissolution observed between experiments conducted in a \qty{0.1}{M} KOH solution and those in \qty{0.4}{M} and \qty{1.0}{M} KOH solutions.

The surface morphology of partially dissolved solid particles provides insights into the rate-controlling mechanisms \cite{Appelo2004,Berner1978}. Particles dissolved under diffusion control typically maintain a smooth surface with a relative absence of selective dissolution and etch patterns, whereas those under chemical reaction control undergo preferential dissolution at high-energy surface regions and exhibit geometrically regular surface features such as etch pits, ledges, and corners. The development of the lamellar structures observed in the dissolved pyrrhotite samples (Fig.~\ref{fig:SEM_Dissolution}) is a clear indicator of preferential dissolution and the involvement of chemical reaction control, agreeing with the interpretations in Sections \ref{sec:temperature} and \ref{sec:kinetic_control_regime}. A closer examination of Fig.~\ref{fig:SEM_Dissolution}a reveals the presence of regions of high surface roughness (marked by arrows) in the initial, cleaned pyrrhotite sample. These regions exhibited distinct striations and contributed to a greater surface area and energy, providing areas where preferential dissolution at higher rates could occur and lead to the formation of the lamellar structures. \textcolor{black}{Similar lamellar structures have also been observed in oxidizing pyrrhotite present in concrete structures affected by iron sulfide reactions \cite{Jana2020,Leemann2023}.}

\begin{figure}[!htb]
	\centering
	\includegraphics[scale=0.9]{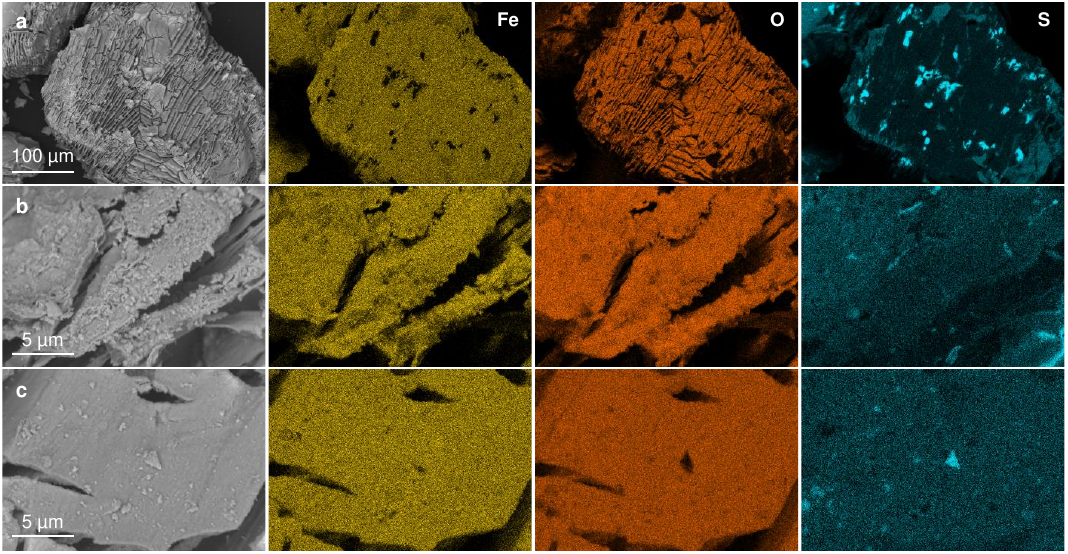}
	\caption{Scanning electron microscopy and elemental mapping images of the pyrrhotite sample after 3-h dissolution in a \qty{0.4}{M} KOH solution: (a) representative area showing lamellar structure and its zoom-in view with (b) rough surface coated with irregularly shaped grains and (c) smoother surface.}
	\label{fig:SEM_EDS}
\end{figure}

To further characterize the chemical composition of the lamellar structures, elemental mapping analysis via EDS was performed on a representative region of the pyrrhotite sample treated with \qty{0.4}{M} KOH. As depicted in Fig.~\ref{fig:SEM_EDS}a, the lamellar structure was primarily composed of Fe and O, suggesting the presence of iron oxide and/or iron (oxy)hydroxide phases as a result of oxidative reactions during dissolution. On the other hand, S was predominantly localized along the edges of the particle and in areas that appeared relatively unaltered. Interestingly, there was an absence of Fe and O in the ``unaltered'' areas (e.g., the ones located in the center of the particle in Fig.~\ref{fig:SEM_EDS}a), indicating a selective dissolution of iron and the formation of elemental sulfur.

The zoom-in views of the lamellar structure in Fig.~\ref{fig:SEM_EDS}a showcase two distinct surface textures. Figure \ref{fig:SEM_EDS}b displays a rough surface coated by irregularly shaped grains, which might indicate the presence of iron (oxy)hydroxide agglomerates according to the elemental analysis as well as the XPS and Mössbauer spectroscopy analyses presented in the following section. In contrast, Fig.~\ref{fig:SEM_EDS}c illustrates a relatively smoother surface texture with no visible sub-micron agglomerates. The elemental analysis suggests that the surface was covered by a thin oxidized layer. The differences in surface texture of the lamellar structure highlight the variability in chemical and physical processes at the micro- and nano-scales within the same sample.

\subsection{Identification of reaction products} \label{sec:products}

The semi-quantitative atomic compositions of the pyrrhotite samples before and after dissolution, derived from the peak areas in XPS survey spectra, are shown in Table \ref{tab:xps_composition}. The initial, cleaned sample exhibited a S/Fe atomic ratio of 2.6, which is significantly higher than the values expected for pyrrhotite. \textcolor{black}{This overestimation of sulfur:metal ratios has also been reported for various transition metal sulfides \cite{Pratt1994,Bonnissel-Gissinger1998,Mycroft1990,Mullet2002,Fuchida2022,Mikhlin2000a}. One possible reason is the greater attenuation of lower kinetic energy photoelectrons from transition metals compared to the higher kinetic energy S(2$p$) photoelectrons, which leads to a systematic undercounting of metal content in XPS measurements. As such, trends rather than absolute values should guide the interpretation of these results.} 

\begin{table}[!htb]
\footnotesize
\centering
\caption{Atomic composition of the pyrrhotite samples before and after 3-h dissolution in different KOH solutions by X-ray photoelectron spectroscopy survey scans.}
\label{tab:xps_composition}
	\begin{tabular}{llllllll}
		\toprule
        Sample    & \multicolumn{4}{l}{Atomic percentage (at.\%)} & \multicolumn{3}{l}{Ratio} \\
        \cmidrule(l){2-8} & K   & O    & S    & Fe   & O/Fe & S/Fe & O/S \\ \midrule
        Initial   & 0.0   & 36.1 & 38.3 & 14.5 & 2.5  & 2.6  & 0.9 \\
        0.1\,M KOH & 1.2 & 50.6 & 23.7 & 14.3 & 3.5  & 1.7  & 2.1 \\
        0.4\,M KOH & 2.4 & 53.0 & 21.5 & 14.9 & 3.6  & 1.4  & 2.5 \\
        1.0\,M KOH & 5.0 & 49.8 & 20.3 & 14.8 & 3.4  & 1.4  & 2.4 \\ \bottomrule
	\end{tabular}
\end{table}

After dissolution, the amount of K in the pyrrhotite samples was found to increase as the KOH concentration in the solutions increased, indicating chemical interactions between the solutions and the samples during the dissolution process. Additionally, there was a notable increase in the O/Fe and O/S ratios of the samples after dissolution, with a corresponding decrease in the S/Fe ratio. These changes suggest that the dissolution process involves substantial oxidation reactions and the release of sulfur into the solution, aligning with the findings discussed in the previous sections. Despite the changes in the O and S contents, the amount of Fe remained relatively consistent across the samples, which is anticipated since any ferrous/ferric ions released during dissolution would likely precipitate due to the highly alkaline environments. 

\begin{figure}[!htb]
	\centering
	\includegraphics[scale=0.9]{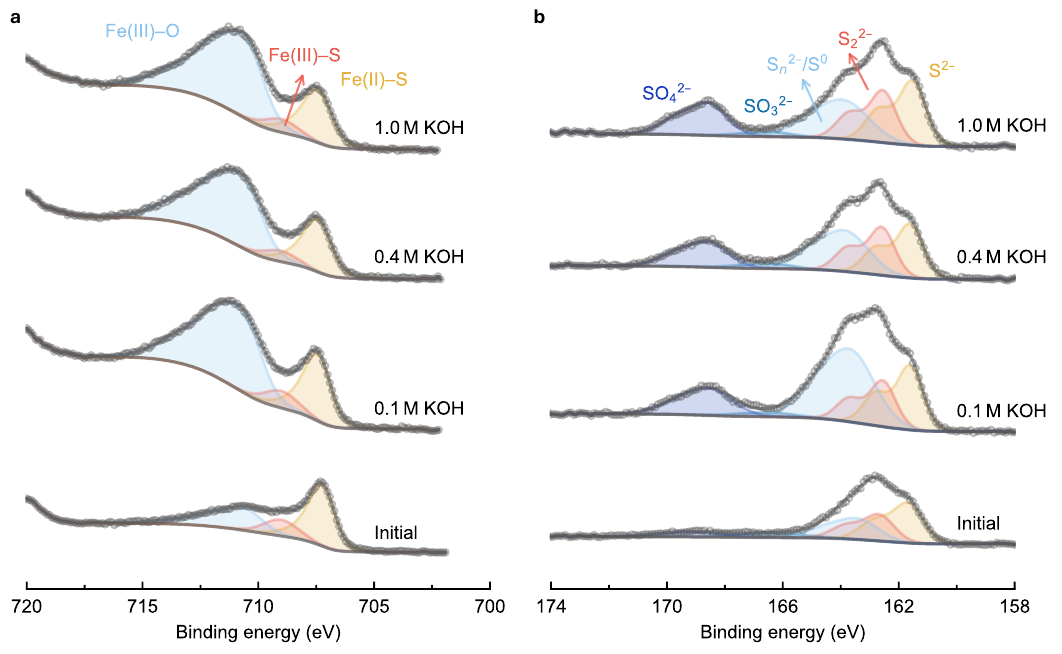}
	\caption{X-ray photoelectron (a) Fe(2$p_{3/2}$) and (b) S(2$p$) spectra of the pyrrhotite samples before and after 3-h dissolution in different KOH solutions by high-resolution scans. Detailed fitting parameters are provided in Table S2.}
	\label{fig:XPS_Species}
\end{figure}

High-resolution Fe(2$p_{3/2}$) and S(2$p$) spectra of the pyrrhotite samples were acquired to extract detailed information about the chemical states and bonding environments ({\it cf.}~Fig.~\ref{fig:XPS_Species}). Detailed information regarding the fitting parameters is provided in Table S2. As shown in Fig.~\ref{fig:XPS_Species}a, the Fe(2$p_{3/2}$) spectrum of the initial sample was dominated by the Fe(II)--S and Fe(III)--S components, located near \qty{707}{eV} and 709--\qty{712}{eV}, respectively. The presence of both components is anticipated in pyrrhotite lattice \cite{Wang2023a} and in good agreement with previous XPS studies on vacuum fractured pyrrhotite \cite{Pratt1994,Mycroft1995}. Despite rigorous efforts to minimize oxidation during sample preparation and storage, there was a minor contribution from the Fe(III)--O component (\num{\sim 711}--\qty{714}{eV}), indicating the high susceptibility of pyrrhotite to air oxidation. After dissolution in KOH solutions, the relative contribution of Fe(III)--O clearly increased, whereas those of Fe(II)--S and Fe(III)--S decreased. These changes suggest the oxidation of S-bonded Fe(II) to O-bonded Fe(III) and the conversion of Fe(III)--S bonds to Fe(III)--O bonds. It is also likely that part of the Fe(III)--S bonds were derived from Fe(II)--S \cite{Wang2023a,Jeong2010}, although further investigation is needed to confirm this reaction pathway. The binding energies for the Fe(III)--O component align closely with those typically observed for goethite \cite{Ferris1989,McIntyre1977} ({\it cf.}~Table S1), implying the formation of Fe(III)-(oxy)hydroxide phases during the dissolution process.

The S(2$p$) spectra of the pyrrhotite samples (Fig.~\ref{fig:XPS_Species}b) reveal the presence of various sulfur species, including monosulfide (\ce{S^2-}) at \qty{\sim 161.6}{eV}, disulfide (\ce{S_2^2-}) at \qty{\sim 162.6}{eV}, polysulfide (\ce{S_n^2-}) or elemental sulfur (\ce{S^0}) at \qty{\sim 163.6}{eV}, sulfite (\ce{SO_3^2-}) at \qty{\sim 166.3}{eV}, and sulfate (\ce{SO_4^2-}) at \qty{\sim 168.6}{eV}. Prior to the dissolution experiments, the pyrrhotite sample mainly consisted of monosulfide, disulfide, and polysulfide/elemental sulfur, similar to the observations in previous studies \cite{Pratt1994,Mycroft1995,Mikhlin2002}. This sulfur configuration originates from S--metal and S--S bonds \cite{Pratt1994,Hyland1989}. Very minor peaks were observed for sulfite and sulfate in the initial sample. After 3-h dissolution, the pyrrhotite samples exhibited a significant increase in the oxy-sulfur species (sulfite and sulfate), highlighting an oxidative transformation of the sulfur species. 

\begin{figure}[!htb]
	\centering
	\includegraphics[scale=0.9]{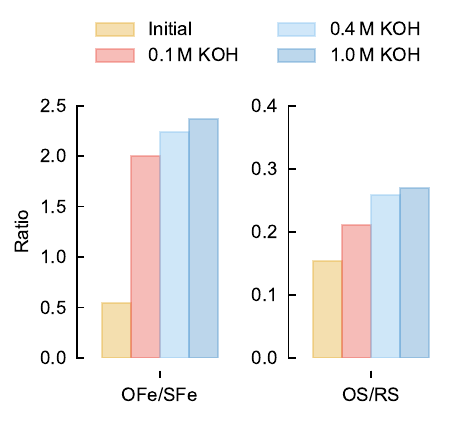}
	\caption{Comparisons of the O-bonded Fe to S-bonded Fe ratio (OFe/SFe) and the oxy-sulfur to reduced sulfur ratio (OS/RS) in the pyrrhotite samples before and after 3-h dissolution in different KOH solutions.}
	\label{fig:XPS_Ratio}
\end{figure}

To analyze the distribution of oxidation states in iron and sulfur within pyrrhotite, two key ratios were quantified: the ratio of O-bonded Fe to S-bound Fe (OFe/SFe) and the ratio of oxy-sulfur to reduced sulfur (including monosulfide, disulfide, and polysulfide/elemental sulfur, with oxidation states ranging from (--II) to (0)) denoted as OS/RS. As depicted in Fig.~\ref{fig:XPS_Ratio}, both the OFe/SFe and the OS/RS ratios increased after the dissolution experiments, with a more pronounced change observed in the OFe/SFe ratio. This trend suggests that substantial oxidation of both iron and sulfur species occurred under the alkaline environments. The less obvious change in the OS/RS ratio is likely due to the release of oxy-sulfur species into the solution, which is not captured by XPS characterization of solid samples. As the KOH concentration increased from \qty{0.1}{M} to \qty{1.0}{M}, an increase in ratios was observed, indicating enhanced oxidation processes at higher pH. This finding agrees well with the observations in solution analyses and surface characterization discussed earlier.

\begin{figure}[!htb]
	\centering
	\includegraphics[scale=0.9]{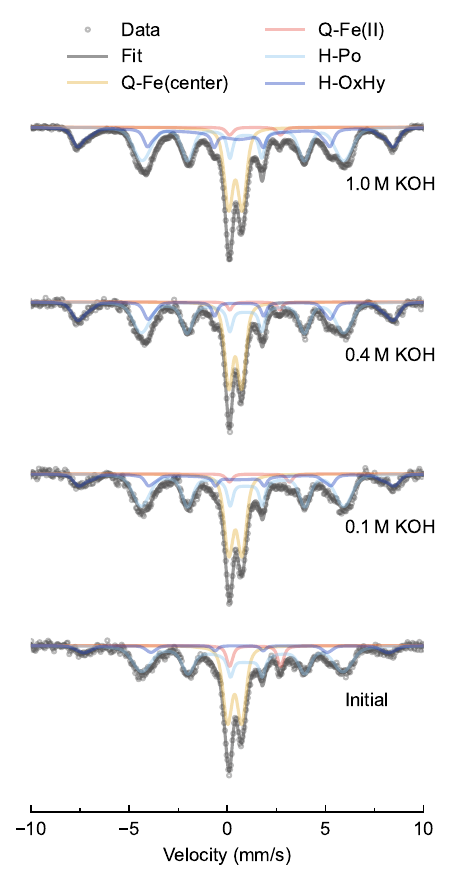}
	\caption{5-K \ce{^57Fe} Mössbauer spectra of the pyrrhotite samples before and after 3-h dissolution in different KOH solutions. The resolved spectra components and assignments are: (1) Q-Fe(center): Fe(III) in silicates, Fe(II) in disordered iron sulfides, or superparamagnetic Fe in pyrrhotite or Fe(III)-(oxy)hydroxide phases; (2) Q-Fe(II): Fe(II) in silicates or adsorbed Fe(II); (3) H-Po: Fe(II) in pyrrhotite; and (4) H-OxHy: Fe(III) in Fe(III)-(oxy)hydroxides, including ferrihydrite and goethite-like phases. Spectra are normalized to relative intensity (y-axis). Detailed fitting parameters are provided in Table S4.}
	\label{fig:Mossbauer_5K_split}
\end{figure}

Fe speciation of the pyrrhotite samples before and after dissolution was further investigated by \ce{^57Fe} Mössbauer spectroscopy. As shown in Fig.~\ref{fig:Mossbauer_5K_split}, four Fe-bearing spectral populations were identified in the 5-K spectra of the samples: (1) a central Fe quadrupole doublet (Q-Fe(center)) corresponding to Fe(III) in silicate impurities (e.g., chlorite, phlogopite, and talc identified by XRD; {\it cf.}~Fig.~\ref{fig:XRD}), low-spin Fe(II) in disordered iron sulfides, or superparamagnetic Fe in pyrrhotite and Fe(III)-(oxy)hydroxide phases that are above their ordering (blocking) temperature; (2) a wide, high-spin Fe(II) quadrupole doublet (Q-Fe(II)) representing paramagnetic Fe(II) in silicate phases or adsorbed Fe(II); (3) a narrow Fe(II) sextet (H-Po) that corresponds to ordered and partially-ordered Fe(II) in pyrrhotite; and (4) a broad sextet (H-OxHy) that corresponds with ordered and partially-ordered Fe(III) in Fe(III)-(oxy)hydroxides, including ferrihydrite and nanocrystalline goethite-like phases. Their relative abundances are summarized in Table \ref{tab:mossbauer_abundance_5K_split} and the detailed Mössbauer spectral parameters are provided in Table S4.

In the initial, cleaned sample, a minor portion of Fe (\qty{\sim 9}{\%}) was magnetically ordered as Fe(III)-(oxy)hydroxides (Table \ref{tab:mossbauer_abundance_5K_split}), which is consistent with the presence of the Fe(III)-O component detected by XPS ({\it cf.}~Fig.~\ref{fig:XPS_Species}). The H-Po component was the most abundant Fe-bearing phase in the sample, accounting for \qty{56}{\%} of total Fe content, followed by Q-Fe(center) for \qty{29}{\%} and Q-Fe(II) for \qty{6}{\%}. After dissolution, a notable reduction in the H-Po and Q-Fe(II) contributions was observed, along with a significant increase in the H-OxHy component. This suggests that Fe(II) within both pyrrhotite and silicate phases were oxidized to Fe(III)-(oxy)hydroxides during the dissolution process. The formation of Fe(III)-(oxy)hydroxide phases identified by Mössbauer spectroscopy aligns closely with the observations by EDS and XPS, as discussed earlier. On the other hand, a slight decrease in the Q-Fe(center) contribution was observed. However, as this component corresponded to multiple potential phases (see Section S2.2 for more discussion), further interpretation remains unclear. 

\begin{table}[!htb]
\footnotesize
\centering
\caption{Spectral areas (\%) of Fe-bearing components in 5-K \ce{^57Fe} Mössbauer spectra of the pyrrhotite samples before and after 3-h dissolution in different KOH solutions.}
\label{tab:mossbauer_abundance_5K_split}
\begin{threeparttable}
	\begin{tabular}{lllll}
		\toprule
    Sample & Q-Fe(center) & Q-Fe(II) & H-Po & H-OxHy \\ \midrule
    Initial & 29.3(10) & 6.20(70) & 55.8(15) & 8.7(14) \\
    0.1 M KOH & 27.4(47) & 2.27(80) & 49.5(84) & 21(10) \\
    0.4 M KOH & 28.2(47) & 1.91(74) & 47.4(81) & 22.4(97) \\
    1.0 M KOH & 24.3(12) & 2.05(24) & 42.2(24) & 31.4(27)\\ \bottomrule
	\end{tabular}
 \begin{tablenotes}
    \item \textit{Note}: Errors in parentheses are given in concise form for the last digit (e.g., 29.3(10) = \num[uncertainty-mode = separate,separate-uncertainty-units=single]{29.3 \pm 1.0}). For component interpretation, see Fig.~\ref{fig:Mossbauer_5K_split}.  
\end{tablenotes}
\end{threeparttable}
\end{table}

\section{Discussion}
\subsection{Dissolution mechanisms of pyrrhotite at high pH}

\begin{figure}[!htb]
	\centering
	\includegraphics[scale=0.9]{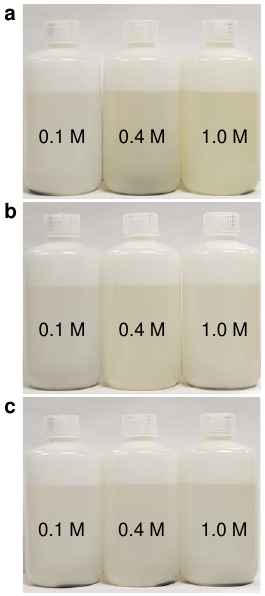}
	\caption{Color changes of different KOH solutions during dissolution experiments at (a) \qty{1}{h}, (b) \qty{2}{h}, and (c) \qty{3}{h}.}
	\label{fig:Bottles}
\end{figure}

The analyses of the released sulfur into aqueous solutions and the solid samples before and after dissolution reveal complex behaviors in pyrrhotite dissolution. As pH increases, a faster sulfur release rate was observed in this study ({\it cf.}~Table \ref{tab:rates}). Dissolved iron was indirectly monitored through the color changes of the solutions at different time intervals during the experiments. As shown in Fig.~\ref{fig:Bottles}, the \qty{0.1}{M} KOH solution remained colorless throughout the experimental duration; however, the color of the \qty{0.4}{M} and \qty{1.0}{M} KOH solutions changed from colorless to pale green and then back to colorless, with the extent and rate of changes dependent on the solution pH. The green color indicates the presence of \ce{Fe^2+} in the solutions. After 1-h dissolution, the \qty{1.0}{M} KOH solution exhibited a light green color, whereas a less pronounced color was observed for the \qty{0.4}{M} KOH solution. At \qty{2}{h}, the color of the \qty{1.0}{M} KOH solution faded, while the \qty{0.4}{M} KOH solution appeared a slightly more intense green color. Both solutions became colorless at the completion of the experiments (\qty{3}{h}). The visible change in color suggests that the fate of iron follows a dissolution--precipitation process. \ce{Fe^2+} is released into the solution during dissolution, especially at higher pH, according to the following reaction \cite{Nicholson1993,Belzile2004}:
\begin{equation}
	\ce{Fe_{1-x}S + (2 - \frac{$x$}{2}) O2 + $x$ H2O -> (1 - $x$) Fe^2+ + SO4^2- + 2$x$ H+} \label{eq:pyrhotite-1}\\ 
\end{equation}
Although no obvious color change was observed in the \qty{0.1}{M} KOH solution, it may be attributed to the very low concentration of \ce{Fe^2+} resulting from the lower extent of the reaction. The oxidation of \ce{Fe^2+} in the aqueous phase occurs quickly at high pH \cite{Millero1985} and produces \ce{Fe^3+} that can precipitate out of the solution as Fe(III)-(oxy)hydroxide phases (e.g., ferrihydrite and goethite):
\begin{gather}
    \ce{Fe^2+ + \frac 14 O2 + H+ -> Fe^3+ + \frac 12 H2O} \label{eq:iron-1} \\
    \ce{Fe^3+ + 3 H2O -> Fe(OH)3 + 3 H+} \\
	\ce{Fe^3+ + 2 H2O -> FeOOH + 3 H+} 
\end{gather}
As evidenced by the solid analyses (EDS, XPS, and Mössbauer spectroscopy) in Sections \ref{sec:surface} and \ref{sec:products}, these Fe(III)-(oxy)hydroxide phases form on the surface of the solid resides and could act as a barrier to diffusion. Along with the apparent activation energy of \qty{43.58}{\kJ\per\mol} derived in Section \ref{sec:temperature} and the analysis by the general shrinking core model in Section \ref{sec:kinetic_control_regime}, these findings indicate that the dissolution of pyrrhotite involves a combination of diffusion through the Fe(III)-(oxy)hydroxide layer and chemical reactions on the surface. The relative contribution of each controlling mechanism is strongly dependent on pH and temperature.

The XPS results ({\it cf.}~Section \ref{sec:products}) provide detailed information on the chemical evolution in the solid samples. Upon dissolution, the Fe(II)--S and Fe(III)--S contributions in the Fe(2$p_{3/2}$) spectra decreased, whereas the Fe(III)--O contribution increased (Fig.~\ref{fig:XPS_Species}a). These changes indicate the occurrence of oxidation of \ce{\bond{#}Fe^{II}-S -> \bond{#}Fe^{III}-O} and \ce{\bond{#}Fe^{III}-S -> \bond{#}Fe^{III}-O} at high pH. Although the reaction pathway of \ce{\bond{#}Fe^{II}-S -> \bond{#}Fe^{III}-S} has been proposed for iron sulfide minerals \cite{Jeong2010}, further research is needed to confirm this mechanism at the highly alkaline conditions.

Various oxidation states of sulfur species were identified in the S(2$p$) spectra (Fig.~\ref{fig:XPS_Species}b). While some sulfur was oxidized to sulfate, the oxidation was not complete and the majority of oxidized sulfur was in intermediate oxidation states, including sulfite, polysulfide/element sulfur, and disulfide. The presence of these intermediate species implies a stepwise oxidation process. Notably, the contribution of the polysulfide/element sulfur component was remarkably high in the pyrrhotite sample after dissolution in a \qty{0.1}{M} KOH solution. This suggests a potential accumulation of polysulfide/element sulfur during dissolution prior to further oxidation into sulfate species. Along with the evidence of the formation of elemental sulfur by EDS (Section \ref{sec:surface}), the incomplete reaction of pyrrhotite at high pH can also be described as \cite{Nicholson1993}:
\begin{equation}
	\ce{Fe_{1-x}S + (\frac{1 - $x$}{2}) O2 + 2(1 - $x$) H+ -> (1 - $x$) Fe^2+ + S^0 + (1 - $x$) H2O}
\end{equation}

\subsection{pH dependence of dissolution rates} \label{sec:pH_dependence}

\begin{figure}[!htb]
	\centering
	\includegraphics[scale=0.9]{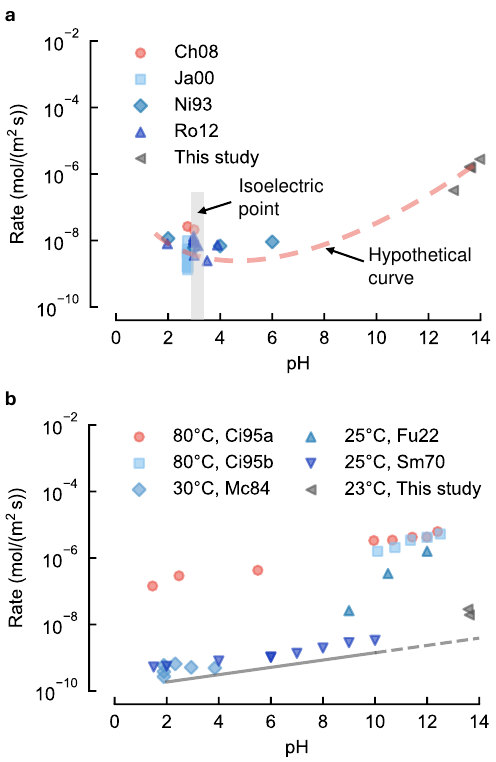}
	\caption{Effect of pH on dissolution rate of (a) pyrrhotite and (b) pyrite. Data presented in (a) are from experiments conducted at 20--\qty{25}{\celsius}. The gray solid line represents the rate law from \cite{Williamson1994}, assuming a dissolved oxygen concentration of \qty{0.3}{mM} (i.e., equilibrium with $P_\text{\ce{O2}}=\qty{0.21}{atm}$), whereas the gray dashed line extends beyond the experimental pH range for the rate law to show its applicability at higher pH. The legend refers to the literature sources listed in Table \ref{tab:Lit_data_source} and all data are provided in Table S5.}
	\label{fig:Lit_Data} 
\end{figure}

In order to demonstrate the influence of pH on the dissolution rates of iron sulfides, the derived results in this study are compared with the rates reported in the literature. The sources of the literature are summarized in Table \ref{tab:Lit_data_source}, and compiled rate data (83 measurements in total, including 48 for pyrrhotite and 35 for pyrite) are provided in Table S5. As shown in Fig.~\ref{fig:Lit_Data}a, the dissolution rates for pyrrhotite derived from this study under a pH of 13--14 are approximately two orders of magnitude higher than those collected from the literature under more acidic conditions (e.g., pH 2--4). This substantial increase highlights the strong pH dependence of pyrrhotite dissolution rates. Likewise, the dissolution rates for pyrite also increase with pH in a similar magnitude, as depicted in Fig.~\ref{fig:Lit_Data}b. Although the dissolution rates for pyrite are generally lower than those for pyrrhotite, the trend of increasing rates with higher pH appears somewhat comparable for both minerals.

\begin{table}[!htb]
\fontsize{9}{10}\selectfont
\centering
\caption{Summary of the literature sources for dissolution rate data of pyrrhotite and pyrite.}
\begin{threeparttable}
\begin{tabular*}{\linewidth}{@{\extracolsep{\fill}}p{3.1cm}lp{1.5cm}lllll}
\toprule
\makecell[tl]{Reference\\ identifier} & pH & Solution & \makecell[tl]{Temperature\\ (\unit{\degreeCelsius})} & \makecell[tl]{Particle \\ size (\unit{\um})} & \makecell[tl]{Reactor\\ type} & \makecell[tl]{Reaction \\progress \\variable} & \makecell[tl]{Number \\of data} \\ \midrule
\textit{Pyrrhotite} &  &  &  &  &  &  &  \\
\hspace{2mm}\makecell[tl]{Ch08 \\(Chiriţǎ {\it et al.}~\cite{Chirita2008})} & 2.75--3.00 & HCl & 25--45 & 149 & BR & Hydrogen ion & 4 \\
\hspace{2mm}\makecell[tl]{Ja00 \\(Janzen {\it et al.}~\cite{Janzen2000})} & 2.75 & HCl & 25 & 45--250 & MFR & Iron & 12 \\
\hspace{2mm}\makecell[tl]{Ni93 \\(Nicholson and \\Scharer \cite{Nicholson1993})} & 2--6 & EDTA, \ce{HNO3}, NaOH & 10--33 & 90--125 & MFR & Iron & 11 \\
\hspace{2mm}\makecell[tl]{Ro12 \\(Romano \cite{Romano2012})} & 1.97--3.90 & Seawater, HCl & 4--35 & 45--150 & BR & Iron & 15 \\
\addlinespace
\textit{Pyrite} &  &  &  &  &  &  &  \\
\hspace{2mm}\makecell[tl]{Ci95a \\(Ciminelli and \\Osseo-Asare \cite{Ciminelli1995})} & 1.5--12.4 & HCl, \ce{Na2CO3} & 80 & 53--106 & BR & Solid mass & 8 \\
\hspace{2mm}\makecell[tl]{Ci95b \\(Ciminelli and \\Osseo-Asare \cite{Ciminelli1995a})} & 10.1--12.5 & NaOH & 80 & 53--106 & BR & Solid mass & 5 \\
\hspace{2mm}\makecell[tl]{Fu22 \\(Fuchida {\it et al.}~\cite{Fuchida2022})} & 9--12 & HCl, NaOH & 25 & 106--150 & BR & Sulfur & 3 \\
\hspace{2mm}\makecell[tl]{Mc84 \\(McKibben \cite{McKibben1984})} & 1.89--3.85 & HCl, NaCl & 20--40 & 125--250 & BR & Iron & 8 \\
\hspace{2mm}\makecell[tl]{Sm70 \\(Smith and \\Shumate \cite{Smith1970})} & 1.5--10.0 & n/r & 25 & 150--250 & MFR & Oxygen & 9 \\ \bottomrule
\end{tabular*} 
\begin{tablenotes}
    \item \textit{Note}: Data were compiled for dissolution in the presence of dissolved oxygen. Reference identifiers are composed of the first two letters of the first author's last name and the last two digits of the publication year (with additional letters for identical identifiers). n/r, not reported; BR, batch reactor; MFR, mixed flow reactor. 
\end{tablenotes}
\end{threeparttable}   
\label{tab:Lit_data_source}
\end{table}

The pH dependence of pyrite dissolution has been well documented  \cite{Smith1970,Williamson1994,Ciminelli1995,Ciminelli1995a}. However, the mechanism behind the increase in dissolution rate at higher pH remains unclear. The aqueous reaction of pyrite during dissolution under atmospheric oxygen can be generally described as \cite{Williamson1994,Descostes2004}:
\begin{equation}
	\ce{FeS2 + \frac{7}{2} O2 + H2O -> Fe^2+ + 2 SO4^2- + 2 H+} \\ 
\end{equation}
Under strongly acidic conditions ($\text{pH}<4$), \ce{Fe^3+} produced through reaction \eqref{eq:iron-1} will largely remain in solution and become an additional oxidant of pyrite \cite{Descostes2004,Williamson2006}:
\begin{equation}
	\ce{FeS2 + 14 Fe^3+ + 8 H2O -> 15 Fe^2+ + 2 SO4^2- + 16 H+} \\ 
\end{equation}
As such, the dissolution of pyrite at low pH involves oxidation reactions by dissolved oxygen and \ce{Fe^3+}. As pH increases from 4 to 8, \ce{Fe^3+} becomes less soluble and tends to precipitate as Fe(III)-(oxy)hydroxide phases on mineral surfaces \cite{Furcas2022,Wieland2023}. Nevertheless, as depicted in Fig.~\ref{fig:Lit_Data}b, the formation of these products does not significantly inhibit the reaction or passivate the surface \cite{Bonnissel-Gissinger1998,Ciminelli1995}, and the reaction rate continues to increase with solution pH. Although a clear explanation of this phenomenon has not been given, Williamson and Rimstidt \cite{Williamson1994} proposed a rate law for pyrite dissolution over the pH range of 2--10 based on a compilation of literature data:
\begin{equation}
r=10^{-8.19} m_{\mathrm{DO}}^{0.5} m_{\ce{H+}}^{-0.11}
\end{equation}
where $r$ is the rate of pyrite destruction (\unit{\mol\per\square\m\per\s}) and $m_{\mathrm{DO}}$ and $m_{\ce{H+}}$ denote the concentrations of dissolved oxygen and proton, respectively. Extrapolation of this rate law to highly alkaline conditions suggests it may reasonably describe the experimental data observed in this study ({\it cf.}~Fig.~\ref{fig:Lit_Data}b). 

Similar to pyrite, pyrrhotite can also be oxidized by \ce{Fe^3+} at low pH range \cite{Nicholson1993,Belzile2004}:
\begin{equation}
	\ce{Fe_{1-x}S + (8 - 2$x$) Fe^3+ + 4 H2O -> (9 - 3$x$) Fe^2+ + SO4^2- + 8 H+}\\ 
\end{equation}
Additionally, proton-promoted dissolution (non-oxidative dissolution) can occur \cite{Belzile2004}: 
\begin{equation}
	\ce{Fe_{1-x}S + 2 H+ -> (1 - $x$) Fe^2+ + H2S}\\ 
\end{equation}
At $\text{pH}< \num{\sim 2}$, the rate of proton-promoted dissolution has been found to be significantly higher than that of oxidative dissolution governed by oxygen or \ce{Fe^3+} \cite{Chirita2014,Thomas1998}, which reasonably explains the behaviors of pyrrhotite dissolution under acidic conditions ({\it cf.}~Fig.~\ref{fig:Lit_Data}a). Nevertheless, the mechanism at higher pH remains largely unexplored in existing literature. Moreover, there is a noticeable gap in experimental data at the circumneutral and mildly alkaline pH ranges, hindering a comprehensive understanding of the dissolution process. Chiriţǎ and Rimstidt \cite{Chirita2014} developed an oxidative dissolution rate law for pyrrhotite as a function of temperature and $P_\text{\ce{O2}}$, using literature data over a pH range of 1.97--3.5. However, the predictive performance of this equation is relatively low (with a correlation coefficient of \qty{21}{\%}). The unsatisfactory performance of the model results from the relatively small amount of data, inconsistent experimental conditions, and uncontrolled variables such as reactor design, mineral impurity, and stirring speed. Future work is necessary to collect internally consistent and extensive data to derive a rate equation for pyrrhotite dissolution across both acidic and alkaline conditions. 

The pH dependence of mineral dissolution rates is often associated with the isoelectric point of the mineral \cite{Harries2013,Crundwell2016a,OsseoAsare1996,Brantley2008}. Here, we explore its applicability to iron sulfide dissolution. In aqueous solutions, pyrrhotite is expected to develop a hydroxyl group (\ce{\bond{#}Fe-OH}) and a thiol group (\ce{\bond{#}S-H}) on its surface \cite{Bebie1998}. These two surface functional groups can undergo protonation or deprotonation, when the solution pH is below or above the isoelectric point, respectively. Both processes result in polarization and weakening of the underlying Fe--S bonds, allowing the detachment of surface atoms into solutions. The isoelectric point of pyrrhotite typically lies in the range of 3.0--3.3 \cite{Bebie1998}. As such, the enhanced dissolution rate of pyrrhotite at pH below 2.70 observed in Harries {\it et al.}~\cite{Harries2013} and in Fig.~\ref{fig:Lit_Data}a can be attributed to the dominance of protonated functional groups (e.g., \ce{\bond{#}S-H2+}) on the mineral surface that weakens the bonding with the underlying lattice (referred to as proton-promoted dissolution). Although experimental data are scarce, pyrrhotite dissolution rates appear to increase with pH above the isoelectric point, as shown in Fig.~\ref{fig:Lit_Data}a. We herein hypothesize that, in addition to the commonly understood oxidative dissolution driven by oxygen (and/or \ce{Fe^3+}), hydroxyl-promoted dissolution may play a critical role at high pH, by favoring the deprotonation of the surface functional groups (\ce{\bond{#}S-H -> \bond{#}S-^-}) and thus weakening the Fe--S bonds. Likewise, the isoelectric point of pyrite falls below 2 \cite{Bebie1998}. The observed increase in pyrite dissolution rates with increasing pH beyond this threshold shown in Fig.~\ref{fig:Lit_Data}b could also be attributed, in part, to deprotonation processes. 

It is also worthwhile to consider the solubility of Fe(III) as another potential thermodynamic driving force for the dissolution of iron sulfides \cite{Nagy1992}. Fe(III)-(oxy)hydroxide phases are more soluble at high or low pH than at circumneutral pH \cite{Furcas2022,Wieland2023}. The higher solubility results in larger deviations from equilibrium upon the dissolution of iron sulfides, potentially contributing to the enhanced rates. This somewhat aligns with the observed trend in pyrrhotite dissolution rates as a function of pH ({\it cf.}~Fig.~\ref{fig:Lit_Data}a). Nevertheless, it may not fully account for the behavior observed in pyrite dissolution, where rates increase from low to high pH (Fig.~\ref{fig:Lit_Data}b). This disparity might be attributed to the combined contributions of multiple mechanisms involved in the dissolution: proton-driven dissolution (controlled by pH and solubility) and redox-driven dissolution (by oxygen and \ce{Fe^3+}). The relative importance of each mechanism likely varies across various pH levels and among different iron sulfide minerals. Further research will be needed to fully elucidate these processes.

\subsection{Implications for iron sulfide reactions in concrete and future work} \label{sec:implications}

Understanding the dissolution mechanisms of iron sulfides in alkaline environments provides valuable insights into the long-term behavior of concrete structures containing these minerals. Results from this study suggest that the dissolution rate of pyrrhotite (the more reactive form of iron sulfides) increases significantly with pH \textcolor{black}{(8.7-fold from pH 13 to 14)}. This finding is of practical importance as the control of pore solution pH could be a potential strategy to slow down the dissolution rate and thus mitigate the deleterious reactions. This can be achieved by the use of low-alkali cements and supplementary cementitious materials \textcolor{black}{\cite{Thomas2011a,Lothenbach2011}. Nevertheless, it remains an important open question whether such strategies merely delay reactions or effectively prevent them within the lifetime of a structure. It is thus essential to establish a fundamental understanding of the nature and amount of reactive iron sulfides in aggregates as well as the maximum pH threshold required to mitigate deterioration for specific aggregates. Moreover, adjusting the binder composition can alter the hydrate assemblage and potentially influence the formation of expansive products and the resulting crystallization pressure \cite{Kunther2013,Kunther2015}, which should be taken into account in future investigations. On the other hand}, the product layer of Fe(III)-(oxy)hydroxides formed on the mineral surface was found to be a diffusion barrier, affecting the dissolution kinetics. Promoting the formation of this layer and/or tailoring its properties in solution \textcolor{black}{(e.g., through chemical admixtures)} thus offer alternative solutions to passivate the surface and inhibit further dissolution. \textcolor{black}{A recent demonstration of the dissolution-inhibiting effect of nitrite-based admixtures \cite{Li2025} underscores the potential of this approach.} 

\textcolor{black}{The strong pH dependence of iron sulfide dissolution rates observed in this study ({\it cf.}~Fig.~\ref{fig:Lit_Data}) also provides important insights into oxidation rates under various real-world scenarios. Oxidation is anticipated to occur most rapidly in uncarbonated concrete due to its highly alkaline environment, slower in carbonated concrete where pH is reduced (as reported in \cite{Leemann2023}), and even slower in stockpiles after quarrying and crushing, where sulfuric acid production during oxidation can create acidic conditions. Nonetheless, differences in oxygen and moisture availability, which are not specifically addressed here, can also affect the overall oxidation rate.}

Although the type of alkali was found to have minimal influence on the dissolution behavior of iron sulfides in this study, the role of other ions---such as calcium, aluminum, iron, magnesium, and sulfate---warrants further investigation. These species are commonly present in aqueous systems, especially concrete pore solutions \cite{Vollpracht2016}. Previous research on mineral dissolution has demonstrated that these ions can affect dissolution rates by modifying near-surface solute/solvent properties and mineral reactivity \cite{Dove1997,Snellings2013,Bagheri2022}. For example, recent studies have indicated that calcium and magnesium can slow down the dissolution rate of pyrite at pH levels between 9 and 12 \cite{Fuchida2022}. Additionally, the presence of hematite or alumina suspensions in deionized water has been observed to suppress or enhance pyrite dissolution, respectively \cite{Tabelin2017}. A deeper understanding of the influence of these ions on iron sulfide dissolution in highly alkaline systems could enable more effective control and mitigation of iron sulfide-induced reactions in concrete.

While this study utilizes natural pyrrhotite and pyrite minerals for investigation, the developed experimental framework is applicable to \textcolor{black}{concrete aggregates} for evaluating their oxidation potential and reactivity. \textcolor{black}{It is important to note that concrete aggregates often contain diverse mineralogical phases and impurities, which may influence iron sulfide dissolution kinetics and reaction pathways in real-world scenarios. Future work addressing these compositional complexities will help bridge the gap between fundamental insights and practical applications.} On the other hand, the solutions used in this study can be further customized and tailored to mimic the pore solution chemistry of various cementitious systems according to specific design requirements. In doing so, this experimental framework offers a more robust and insightful alternative to existing chemical testing approaches \cite{Chinchon-Paya2012,Ramos2016,Guirguis2017}, which rely on water or oxidizing solutions (such as \ce{H2O2} or NaOCl) that deviate from typical concrete environments\textcolor{black}{, and provides a pathway to establish limit values for iron sulfides in concrete aggregates}.

The characterizations and analyses presented in this study primarily focus on bulk chemistry. Although they provide substantial insights, detailed information on near-surface evolution during dissolution warrants further research. Additionally, the compilation of pyrrhotite and pyrite dissolution rates (Fig.~\ref{fig:Lit_Data}) reveals data gaps within the circumneutral and alkaline pH ranges, particularly a notable absence of rate data between pH 6 and 13. Most existing experimental data fall within lower pH conditions. Although this work, to our knowledge, presents the first investigation of iron sulfide dissolution in highly alkaline solutions, more experiments are needed to develop comprehensive rate equations that account for the influence of a broader range of pH and temperature. These equations will be crucial for quantitatively understanding the reactivity of iron sulfides in diverse reaction conditions relevant to engineering applications, such as service life predictions of concrete containing iron sulfide-bearing aggregates.

\section{Conclusions}

This study investigates the kinetics and mechanisms of iron sulfide dissolution in alkaline solutions. Dissolution rates were derived from initial slopes of released sulfur concentration versus time. The roles of \textcolor{black}{iron sulfide type}, alkali cation, pH, and temperature were analyzed and discussed. Experimental characterization and kinetic modeling were coupled to reveal the dissolution behavior of iron sulfide minerals. The main findings from this study are summarized as follows:

\begin{itemize}

\item Pyrrhotite exhibited a much higher dissolution rate (approximately three orders of magnitude) and greater reactivity in highly alkaline solutions than pyrite. The behavior of these minerals remained consistent regardless of the alkali type (potassium or sodium) in the solution.

\item Pyrrhotite dissolution rates increased significantly with pH and temperature. An increase in pH from 13 to 13.6 and 14.0 resulted in 5.1- and 8.7-fold accelerations in the rate, respectively, while an increase in temperature from \qty{23}{\celsius} to \qty{40}{\celsius} and \qty{60}{\celsius} resulted in 2.4- and 7.1-fold accelerations. The apparent activation energy of pyrrhotite dissolution at pH 13.6 was derived to be \qty{43.58}{\kJ\per\mol}.

\item The fate of iron was found to follow a dissolution--precipitation process. After the dissolution experiments, extensive oxidation of iron species was observed in the solid residues using both XPS and \ce{^57Fe} Mössbauer spectroscopy, and the oxidation of sulfur species was specifically identified through XPS. The reacted pyrrhotite particles displayed lamellar structures predominantly composed of Fe(III)-(oxy)hydroxides, possibly resulting from preferential dissolution in regions of excess surface roughness and energy.

\item The kinetic control regime of pyrrhotite dissolution was determined by the general shrinking core model and found to be controlled by a combination of surface chemical reactions (oxidation of iron and sulfur species) and diffusion (mainly through a product layer of Fe(III)-(oxy)hydroxides). The relative contribution of each controlling mechanism varied significantly with the solution pH and temperature.

\item A potential mitigation strategy to control iron sulfide reactions in concrete involves reducing the pH of concrete pore solutions. This can be effectively achieved by using low-alkali cements and supplementary cementitious materials\textcolor{black}{, although further research is needed to identify the pH thresholds required to prevent deleterious expansion}. Additionally, engineering the Fe(III)-(oxy)hydroxide layer developed on the mineral surface with specific admixtures could provide an alternative approach to passivate the minerals and inhibit their dissolution and further reactions.

\end{itemize}

\section*{Acknowledgements}

The authors would like to thank Dr.~Kwadwo Osseo-Asare for many enlightening discussions. Laura J. Liermann, Nichole M. Wonderling, Julie M. Anderson, Gino Tambourine, and Nehru S. Mantripragada are acknowledged for their help with total sulfur content and ICP-AES, XRD, SEM/EDS, BET surface area, and Mössbauer spectroscopy analyses, respectively. Angelica Hunt is acknowledged for assistance in developing the dissolution test. This research was supported by the U.S. National Science Foundation (NSF) under Award CMMI \#1825386. Any opinions, findings and conclusions or recommendations expressed in this material are those of the authors and do not necessarily reflect the views of the NSF.

\bibliographystyle{elsarticle-num}
\biboptions{sort&compress}

\end{document}